# Microscale investigation of binary droplet coalescence using a microfluidic hydrodynamic trap

*Shweta Narayan[1], Iaroslav Makhnenko[1], Davis B. Moravec[2], Brad G. Hauser[2], Andrew J. Dallas[2] and Cari S. Dutcher[1§]*

[1]*Department of Mechanical Engineering, University of Minnesota – Twin Cities, Minneapolis, MN*
[2]*Donaldson Company, Inc., Bloomington, MN*
[§]*Corresponding author email: cdutcher@umn.edu*

**Abstract**

Coalescence of micrometer-scale droplets is impacted by several parameters, including droplet size, viscosities of the two phases, droplet velocity and angle of approach, as well as interfacial tension and surfactant coverage. The dynamics and thinning of films between coalescing droplets can be particularly complex in the presence of surfactants, due to the generation of Marangoni stresses and reduced film mobility. In this work, a microfluidic hydrodynamic "Stokes" trap is used to gently steer and trap surfactant-laden micrometer-sized droplets at the center of a cross-slot. Incoming droplets are made to coalesce with the trapped droplet, yielding measurements of the film drainage time. Water droplets are formed upstream using a microfluidic T-junction, in heavy and light mineral oils and stabilized using SPAN 80, an oil-soluble surfactant. Film drainage times are measured as a function of continuous phase viscosity, incoming droplet speed, trapped droplet size, and surfactant concentrations above and below the critical micelle concentration (CMC). As expected, systems with higher surfactant concentrations, higher continuous phase viscosity, and slower droplet speed exhibit longer film drainage times. Perhaps more surprisingly, larger droplets and high confinement also result in longer film drainage times. The results are used here to determine critical conditions for coalescence, including both an upper and a lower critical Capillary number. Moreover, it is shown that induced surfactant concentration gradient effects enable coalescence events after the droplets had originally flocculated, at surfactant concentrations above the CMC. The microfluidic hydrodynamic trap provides new insights into the role of surfactants in film drainage and opens avenues for controlled coalescence studies at micrometer length scales and millisecond time scales.

# 1. Introduction

Droplet coalescence is a fundamental physical phenomenon driving the separation of liquid-liquid systems or emulsions. In the context of emulsion separation, it refers to the combination of dispersed liquid droplets to form larger drops, which can then be more easily removed from the bulk liquid phase. An example of one such application is the removal of dispersed water droplets from diesel fuel, often encountered in automobile fuel delivery systems. Water that enters diesel fuel lines is considered a contaminant due to its tendency to cause microbial growth, rust and corrosion of parts such as common rail injectors.[1] Often, these micrometer-sized dispersed water droplets can be challenging for filter media to remove due to surface-active additives present in the fuel. Examples of surface-active fuel additives include polyisobutylene succinimides (PIBSI), a deposit control additive, and monoolein, a lipid in biodiesel.[2,3] These surface-active additives, or surfactants, stabilize the water-in-fuel emulsion by lowering the interfacial tension, thereby leading to inhibition of coalescence of microscale droplets. Similarly, surfactant-stabilized oil-in-water emulsions can be formed onboard ships, resulting in bilge-water that is extremely difficult to purify before being discharged into the ocean.[4] To facilitate better separation strategies for surfactant-stabilized emulsions, it becomes essential to understand fundamentally the process of droplet-droplet coalescence under well-controlled flow fields on relevant length and time scales.

Emulsions are destabilized through various processes including coalescence, Ostwald ripening, creaming or sedimentation of the dispersed phase.[5] Bulk, or large-scale techniques, study the overall stability of emulsions over time with or without flow present, and several simple but useful techniques have been developed by researchers for this purpose. An example of a bulk technique is vial tests, where the stability of pre-prepared emulsions with varying surfactant concentration, dispersed fraction or salinity is monitored over a specified period. The change in droplet size can be measured by drawing samples at specific time points, and generally, a shift in size distribution towards larger droplet sizes is observed as the emulsion is destabilized.[4,6] Another bulk technique is stir tanks placed in a temperature-regulating bath, where the size distribution is monitored visually or through laser diffraction particle sizing.[7,8] Further, Hudson et al. developed a shear cell consisting of a parallel-plate geometry,[9] to measure coalescence efficiency under shear flow, as well as interfacial tension, using the drop retraction technique.[10] Bulk techniques provide a wealth of information about the collective behavior of droplets in an emulsion, but it is often difficult to tease out the contribution of individual processes leading to emulsion destabilization, such as coalescence or sedimentation. For this reason, microscale techniques have been employed in recent years to study droplet coalescence in emulsions.

Microfluidic techniques have been widely employed for studying multiple-droplet coalescence.[11] One such device design by Dudek et al.[12] utilizes a T-junction geometry to form droplets, followed by a widening in the channel dimensions leading to slowing down and coalescence of droplets. A similar setup by Lin et al.[13] examined the stability of asphaltene-stabilized water in oil emulsions, finding that the presence of demulsifiers and their concentrations had a significant impact on droplet coalescence rates. A microfluidic technique developed by Zhou et al.[14] consists of opposing T-junction geometries where water-in-oil Pickering emulsion droplets are formed, and made to collide in a cross-slot to measure the film drainage time. In addition to multiple-droplet coalescence, it can be valuable to employ controlled microscale single-droplet techniques, discussed in the following paragraphs, to study the fundamental physics of the coalescence process in surfactant-stabilized emulsions.

G. I. Taylor first developed a four-roll mill to study the behavior of emulsions under well-controlled extensional flow fields.[15] This idea was adapted by Leal and coworkers to create a miniaturized and computer-controlled four-roll mill to study droplet deformation and coalescence events at the microscale.[16–19] With this setup, a single drop of a dispersed phase, held stationary at a stagnation point in an extensional flow field, is split into two drops. Subsequently, the flow is reversed, leading to coalescence of the dispersed drops, and a camera captures images of the coalescing drops, yielding measurements of the film drainage time.[16] This device is capable of studying droplet coalescence under varying flow conditions, viscosity ratios, surfactant concentrations and with controlled droplet trajectories.[17] Another controlled coalescence technique, called the 'Cantilevered Capillary Force Apparatus' or CCFA involves the use of micro-capillaries in a flow cell, and has been used by Frostad et al.[20,21] to measure the film drainage time during coalescence of microscale droplets compressed under a constant force. Micro-capillaries and atomic-force microscopy (AFM)-inspired setups have gained tremendous popularity for studying coalescence of microscale droplets in a controlled fashion.[22,23]

The experimental techniques discussed above have offered useful insights into the coalescence behavior of droplets under various physicochemical conditions, therefore aiding in the development of theories and predictive models of coalescence. Coalescence of droplets is known to occur in three distinct stages – the first stage is the approach of droplets under the influence of external flow, the second is the drainage of the thin film of liquid between droplets in contact, and the third and final stage is the rupture of the thin film driven by attractive short-range van der Waals forces leading to coalescence. The drainage of the film between coalescing droplets is often considered to be the pivotal step in droplet coalescence, with the literature focusing on the effect of variables such as viscosity ratio, speed and surfactant concentration on film drainage time. Ivanov[24] and Chesters[25] have both made invaluable contributions to our current understanding of the stages of droplet coalescence. Chesters[25,26] divided the problem into an outer problem, concerning the external flow field and global drop deformation, and an inner problem, concerning the thin film dynamics during drainage. A common theme in the theoretical treatment of coalescence by both Ivanov and Chesters is the classification of the thin film as being immobile, partially mobile or mobile. This classification pertains specifically to the boundary condition at the liquid-liquid interface within the thin film, with immobile referring to the occurrence of a no-slip condition at the interface, mobile referring to a no-stress or slip condition at the interface, and partially mobile referring to a superposition of a no-slip and a uniform flow condition at the interface.

During coalescence, while fluid viscosities, flow type and strength are known to play important roles, the influence of surfactants on each of the stages of coalescence is not clearly understood from a fundamental standpoint. Obviously, an increase in surfactant or emulsifier concentration leads to an increase in emulsion stability, and in addition to a reduction in interfacial tension, steric hindrance by surfactant molecules or electrostatic interactions may enhance the interface's ability to resist coalescence.[24] Nevertheless, it can be tricky to pin-point the specific mechanism at play during a specific stage of coalescence. For instance, Hudson et al. postulated that surfactants may alter the trajectory of droplet motion during the droplet approach stage by retarding interfacial motion, and that this may lead to suppression of coalescence.[9] Moreover, several studies have shown that surfactant transport to curved interfaces is governed not only by the surfactant concentration in the bulk phase, but also by the curvature or radius of the interface.[27] With water-in-diesel fuel emulsions, we previously showed that diffusion of surfactant to a highly curved, micrometer-sized droplet in flow is faster than that to a planar interface, resulting in the transport shifting towards being limited by kinetics rather than diffusion.[28] These observations lead to the conclusion that coalescence behavior in emulsions, which contain micrometer-scale droplets, could be influenced not only by the surfactant concentration, but also by the surfactant transport to and at the interface. Therefore, there is clearly a pressing need to conduct more controlled single-droplet coalescence experiments in the presence of surfactants on length and time scales relevant to real emulsions with microscale dispersed droplets.

In this work, we employ the microfluidic hydrodynamic "Stokes trap" developed by Shenoy et al.[29] to gain fundamental insights into coalescence of water-in-oil emulsions stabilized by surfactants. The Stokes trap is a four or six channel cross-slot device which provides precise control over the positions of droplets by adjusting the flow rates in the intersecting channels to gently steer droplets using one or more stagnation points. Previously, a circular-shaped six-channel cross-slot was employed by Kumar et al. to perform coalescence of highly confined water-in-oil droplets stabilized by SDS.[30] In the present work the four-channel Stokes trap setup by Shenoy et al.[29] was modified to introduce a T-junction dispersed phase inlet upstream of the trapping region in the four-channel cross-slot.[31] Micrometer-scale water droplets are generated in light and heavy mineral oils containing a range of concentrations of the surfactant SPAN 80. Incoming droplets are allowed to interact or coalesce with trapped droplets in a four-channel Stokes trap.

The device design and setup are detailed in Section 2. Section 3 elaborates on the measurement of film drainage time and its importance as a metric for characterizing droplet coalescence. Film drainage times for water droplets in mineral oils with four different concentrations of SPAN 80 are measured, and correlated with droplet size, speed, viscosity ratio and SPAN 80 concentration. A range of incoming droplet speeds from 10 µm/s to 500 µm/s can be achieved using the hydrodynamic trap, depending on the systems studied. In a typical application, e.g. an automotive fuel filter, the droplet speeds may vary from 250 µm/s to 600 µm/s. Furthermore, critical conditions for coalescence and flocculation are identified in Section 3, ending with a discussion on the possible role of surfactant-gradient induced Marangoni forces in stabilizing water droplets against coalescence.

## 2. Methods and Materials
*2.1 Methods*

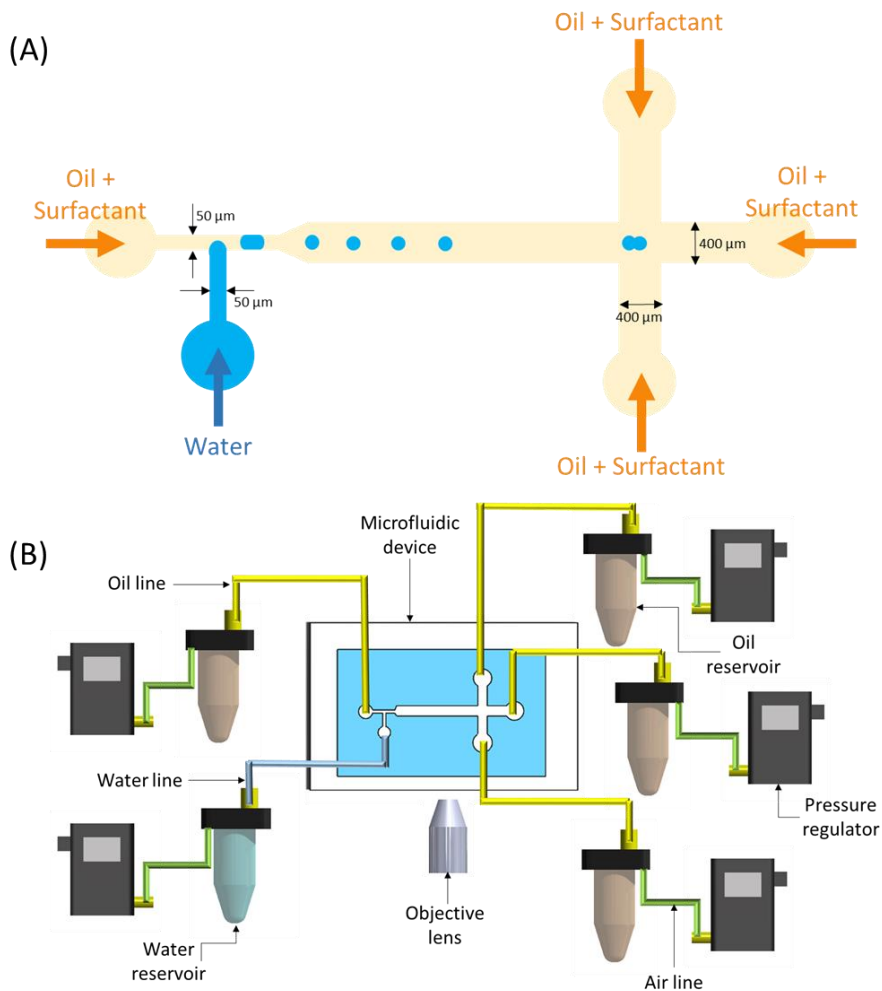

*Figure 1: (A) Schematic of the microfluidic hydrodynamic trap for droplet coalescence studies, showing the T-junction for droplet formation and the cross-slot where droplets are trapped and coalesced. The figure is not drawn to scale. (B) Setup for the hydrodynamic trap showing the device on the microscope stage, along with pressure regulators, fluid supply lines and fluid reservoirs.*

Microfluidic devices are designed using DraftSight 2D CAD software and printed on transparencies with a resolution of 8 µm (CAD/Art Services). The master silicon wafers with the device patterns are then fabricated in a clean room using standard soft lithography techniques[32–34] using SU-8 2050 photoresist (Microchem). Poly (dimethyl siloxane) (PDMS, Sylgard 184, Dow) is poured on the master wafer and peeled off after curing overnight at 70°C. The cured PDMS devices are then cut, holes are punched at the injection ports with a Miltex 1.5 mm biopsy punch, and the devices sealed to glass slides after plasma treatment (Harrick Plasma). The devices are treated using NOVEC 1720 (3M) to render them hydrophobic and soaked overnight in oil. Fluids used in the experiment are stored in fluidic reservoirs (Darwin Microfluidics) and supplied to the device using fluoropolymer Teflon tubing (1/16" OD, 0.02" ID, IDEX Health and Science) connected to a short section of high pressure drop PEEKsil tubing (1/16" OD, 100 µm ID, IDEX Health and Science) using Delrin Union connectors (1/4-28 port, IDEX Health and Science). The high pressure drop tubing is inserted directly into the ports in the device and is intended to prevent backflow and damp flow fluctuations during fluid supply to the microfluidic device. The headspace of air in the fluidic reservoirs is pressurized using analog-controlled pressure regulators (QPV series, Proportion-Air), controlled by voltage signals applied using a National Instruments Data Acquisition (DAQ) board (cDAQ-9174 chassis, NI 9264 AO, NI 9201 AI).

The design of the microfluidic device used for droplet coalescence experiments is adapted from the Stokes trap developed by Shenoy et al.[29] for trapping and manipulating particles in a well-controlled extensional flow field. The Stokes trap is a cross-slot microfluidic device consisting of either four or six intersecting channels, at the center of which either one or two particles can be trapped respectively, by controlling the flow rates in each of the channels. This device has been modified previously by the authors to include a T-junction for controlled droplet formation upstream of the cross-slot.[31] Droplets of water are formed in the T-junction with a shearing phase of mineral oil as shown in **Figure 1**. Droplet size and formation rate can be tuned using a droplet-on-demand program in LabVIEW, which applies short pulses of pressure while also maintaining the liquid-liquid meniscus at the T-junction at all times during the experiment. Downstream of the droplet formation zone, the flow rates in the four arms of the cross slot are adjusted to gently steer a droplet to the center of the cross-slot. Incoming droplets are allowed to coalesce with the trapped droplet.

The flow control strategy employs a Model Predictive Control (MPC) algorithm developed by Shenoy et al., implemented using the Automatic Control and Dynamic Optimization (ACADO) toolkit.[29,35,36] When a droplet enters the region of interest in the cross-slot of the device, the droplet's position is tracked using binary image thresholding using LabVIEW. To steer the droplet to the desired position, here the center of the cross-slot, an objective function which minimizes the distance between the current and set position with small steps in flow rate is solved. With known channel resistances, the calculated flow rates are converted to pressures and applied to the pressure regulators using the DAQ toolbox in LabVIEW (National Instruments). It is possible to control the accuracy of trapping by adjusting the control parameters, including compensatory parameters for deviation from the set position and large changes in flow rate.[35] Videos of coalescing droplets are recorded using a Basler ace acA1300-60gm camera at a frame rate of 60 fps.

Videos of coalescing droplets are analyzed using a custom particle-tracking code in MATLAB. A particle tracking approach by Crocker et al.[37] is integrated into the image analysis code as a function to track droplet positions, and is used to select the desired droplets for measurement of film drainage time. The radius, position, and the instantaneous velocity of each droplet are obtained through image analysis. Additionally, the angle of approach of the incoming droplet with the trapped droplet is defined as angle between the flow direction of the approaching droplet and the line passing through the centers of the two droplets. A graphical user interface in MATLAB is used to track droplets and number them individually. Once the trapped and incoming droplets are defined, the distance between their centers and angle of approach are calculated and tracked until the droplets coalesce. The droplets are assumed to be in contact on the observable length scale when the difference between sum of their radii and distance between centers is less than 1.5% to obtain film drainage time measurements.

*2.2 Materials*

Light and heavy mineral oils (Sigma Aldrich) constitute the continuous fluid phase, and HPLC grade water (Fisher) is the dispersed fluid phase. An oil-soluble surfactant, SPAN 80 (sorbitan monooleate, Croda, Inc.), with an HLB (hydrophilic-lipophilic balance) value of 4.3 is added to the mineral oils at concentrations ranging from 50 – 1000 ppm by volume (0.005% – 0.1% v/v). The interfacial tension and critical micelle concentration (CMC) of SPAN 80 in heavy and light mineral oils is measured using pendant drop tensiometry with a Drop Shape Analyzer (Krüss GmbH) at 22°C using standard procedures [38]. The interfacial tension results from pendant drop tensiometry indicate that the CMC is between 100 ppm by volume (0.01% v/v) and 500 ppm by volume (0.05% v/v) for SPAN 80 in both light and heavy mineral oils as shown in **Figure 2**. Note that since SPAN 80 is oil-soluble, inverse micelles would be formed at concentrations above CMC. The viscosities of light and heavy mineral oils are measured using an AR-G2 rotational rheometer (TA Instruments) over a shear rate range of 1 – 1000 s$^{-1}$. The measurements show that light and heavy mineral oils exhibit Newtonian behavior over the entire range of shear rates, with dynamic viscosities ($\eta_c$) of 26.8 mPa-s and 132.2 mPa-s respectively.

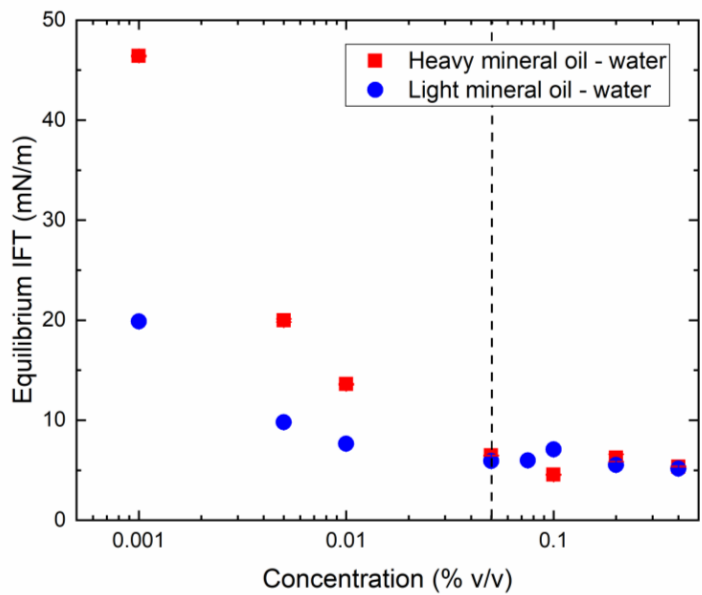

*Figure 2: Pendant drop tensiometry measurements for water drops in light (blue) and heavy (red) mineral oils with SPAN 80 as a surfactant, over a concentration range of 0.001% v/v to 0.4% v/v. The critical micelle concentration (CMC) for both systems is indicated by the black dotted line.*

## 3. Results and Discussion

*3.1 Droplet collision and coalescence: measurement of film drainage time*

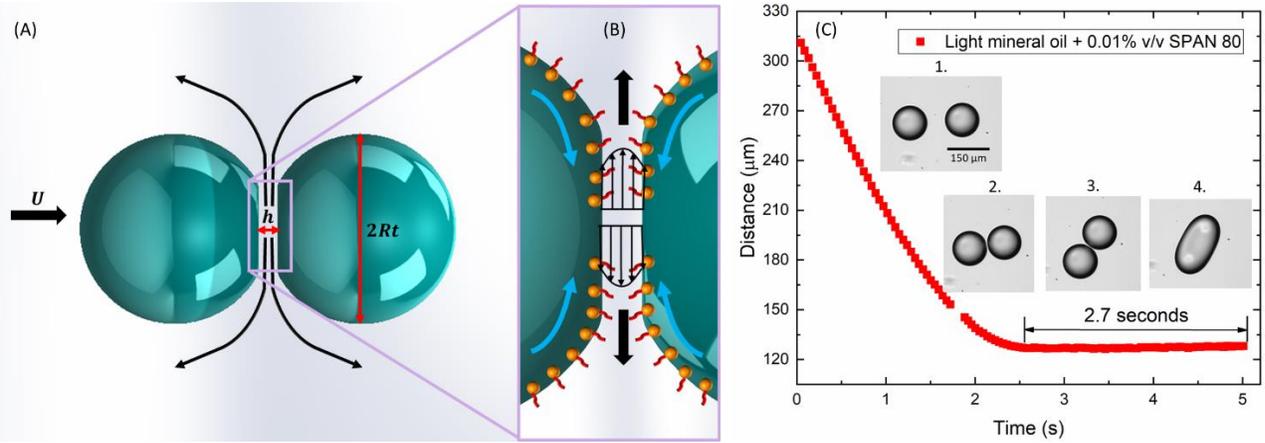

*Figure 3: (A) Schematic showing approach of droplets and formation of a thin film in the hydrodynamic trap. Dimensions of the film are exaggerated for clarity. The incoming droplet velocity, trapped droplet radius and the thin film thickness are indicated in the figure. Black arrows indicate the direction of film drainage flow. (B) Zoomed-in schematic of the thin film formed between coalescing droplets. A partial-slip condition occurs due to presence of surfactants. Surfactants are swept out of the thin film region due to drainage flow, setting up a concentration-gradient driven Marangoni flow indicated by the black arrows. (C) Example of center-to-center distance between droplets as a function of time for droplets of radius ~64 μm, for the light mineral oil + 0.01% v/v SPAN 80 system. The flat portion of the distance versus time plot indicates film drainage. Inset images 1-4 show droplet approach, contact, rotation and coalescence respectively.*

**Figure 3** is a schematic representation showing the coalescence of droplets in an emulsion stabilized by surfactants. In the experiments in this work, the surfactant is soluble in the continuous oil phase. The four-channel microfluidic trap device creates a planar extensional flow field in the cross-slot. In this type of flow, the trapped droplet is located at the center of the cross-slot at the stagnation point, whereas the incoming droplets flow along the compression axis of the flow. Therefore, under the influence of the compressional flow, the incoming droplets approach the trapped droplet at the stagnation point. This approach of the droplets is the first stage in the coalescence process. Once the droplets come into contact, there is a finite time period before the coalescence event occurs. This stage, referred to henceforth as film drainage is marked by the following four features: (1) The distance between the centers of the droplets remains constant on the observable length scale as indicated by the flat region of the distance versus time curve in **Figure 3C**. Note that the size of the thin film is exaggerated in the figure, and in reality, the thickness of the thin liquid film between the droplets is on the nanometer length-scale,[39] which cannot be observed using our microscopy setup; (2) The drops may rotate about each other due to the planar extensional flow field in the cross-slot, as shown in inset image 3 in **Figure 3C**; (3) The thin film of the continuous phase fluid between the droplets drains, although the thinning of the film cannot be visualized on a super-micrometer scale. The film drainage time is the key measurable in our coalescence experiments and is known to be governed by several factors including droplet size, speed, flow type, surfactant type and concentration as well as continuous and droplet phase viscosities.[17] (4) Finally, when the film between the droplets has thinned to the point where attractive van der Waals force becomes dominant, an instability occurs at the droplet interface leading to film rupture and coalescence.[39]

The film drainage time is known to be strongly influenced by the mobility of the droplet interface.[24] As discussed in the introduction, the term mobility refers to the boundary conditions that may exist at the droplet interface in the thin film, namely no-slip, no-stress or partial slip. The mobility is described in coalescence literature as being influenced by two key factors – the viscosity ratio between the droplet and the continuous phase and the surfactant concentration at the interface.[25] In general, systems where the droplet phase is more viscous than the continuous phase, or viscosity ratio $\lambda = \frac{\eta_d}{\eta_c} \gg 1$, where $\eta_d$ and $\eta_c$ are droplet and continuous phase viscosities, are considered to have immobile interfaces. On the other hand, systems where $\lambda \ll 1$ are said to have mobile interfaces. While these definitions are useful from a modeling perspective, many real systems have intermediate values of $\lambda$, also known as partially mobile interfaces. Another definition of the mobility is based on the surfactant concentration in the system. Systems with high surfactant coverage are said to have immobile interfaces, while those with low surfactant coverage are called mobile interfaces. These definitions do not account for the presence of Marangoni stress, which may be dominant in a system depending on the surfactant type and concentration. Again, most real systems are likely to be partially mobile, with the mobility potentially evolving during the film drainage process.[9,40] A more detailed discussion of surfactant effects and mobility is provided in Section 3.4.

Here, we commence our discussion of binary droplet coalescence behavior by first considering the conditions under which droplets do not coalesce. These conditions are referred to as critical conditions for coalescence, and the most common dimensionless number to quantify them is the Capillary number, which is the ratio of viscous to interfacial tension force. It is defined as $Ca = \frac{\eta_c U}{\gamma}$, where $\eta_c$ is the viscosity of the continuous phase, $U$ is the speed of the incoming droplet (defined here as the speed of the incoming droplet at a distance of $4R_t$ from the center of the trapped droplet, where $R_t$ is the radius of the trapped droplet) and $\gamma$ is the equilibrium interfacial tension between the oil and water at a particular SPAN 80 concentration. The distance $4R_t$ is chosen to capture the incoming droplet's velocity before it interacts with the trapped droplet. In the planar extensional flow in the microfluidic hydrodynamic trap, the thin film between droplets must drain for the droplets to coalesce, and this film drainage must occur within the time frame when the planar extensional flow goes from pushing the droplets together to pulling them apart. Thus, there are two important timescales – $t_r$, the residence time of the moving droplet in the cross slot, and $t_d$, the film drainage time. At a critically high $Ca$, $t_r \ll t_d$, and the film between droplets does not drain quickly enough for the droplets to coalesce. In this scenario, the droplets rotate around each other in the flow, the incoming droplet enters the extensional axis of the flow field, and coalescence does not occur.

While this study focuses on droplet coalescence, the critical conditions where coalescence does not occur was also observed for some systems. **Figure 4A** shows one such example for the light mineral oil-0.01% v/v SPAN 80 system. For an incoming droplet speed of ~230.7 µm/s, no coalescence was observed, as the incoming droplet merely rotates around the trapped droplet at the center of the cross-slot and eventually enters the extensional flow quadrant of the device and is advected away from the trapped droplet. It is evident that the droplets were indeed in contact, since the distance between the droplet centers remains constant and equal to the sum of the droplet radii as the droplets rotate around each other. Therefore, the film is unable to drain to the critical film thickness $h_c$ for film rupture within the short period of time ($t_r$ ~ 1.5 seconds) for which the droplets are in contact. The capillary number for this case is calculated to be ~ $1 \times 10^{-3}$, which is the upper limit of $Ca$ for the light mineral oil-0.01% v/v SPAN 80 system, above which coalescence does not occur. While the existence of a upper critical Ca is expected for coalescence in all systems, the actual transition value will also be dependent on surfactant concentration and fluid properties, and $1 \times 10^{-3}$ should not be considered universal.

Interestingly, in addition to an upper limit on the capillary number, it was also observed that when the incoming droplet speed is too small and the surfactant concentration is high (above CMC), the droplets do not coalesce. Instead, the droplets stick together or 'flocculate' and the film between the droplets does not rupture, as shown in **Figure 4B**. We hypothesize that this flocculation would occur in systems laden with high concentrations of surfactant, here 0.1% v/v SPAN 80. In fact, in the heavy mineral oil + 0.1% v/v SPAN 80 system, no coalescence was observed in the experiments, while several flocculation events were observed. Ivanov[24] described the film drainage process in the presence of surfactants as being influenced by interfacial dilatation, bending, electrostatic and ionic interactions as well as steric repulsion. Surfactants lower the interfacial energy by occupying sites at the liquid-liquid interface. The low interfacial energies coupled with steric and other interactions are known to stabilize emulsion systems. Moreover, film drainage may sweep the surfactants out of the thin film, causing strong Marangoni force along the interface directed inwards into the thin film region, which in turn resists film drainage.[41] In the absence of a strong external force leading to film drainage, which occurs at higher velocities of droplet impact, these stresses caused by surfactants at the interface dominate. Additionally, when the incoming droplet has rotated to 90° in the flow, the extensional flow is not strong enough to pull the droplets apart, and the attractive van der Waals force between droplets balances the force due to the extensional external flow. Thus, flocculation of droplets indicates a balance between the external force and the interactive forces between droplets, which prevents the film from rupturing. One such example is shown in **Figure 4B** for the light mineral oil – 0.1% v/v SPAN 80 system, where the surfactant concentration is above the CMC. In this case, the droplets flocculate or remain stuck together without coalescing. The capillary number in this case is $Ca \sim 1 \times 10^{-4}$, and represents the lower limit of $Ca$ for this system, below which the droplets flocculate. Such flocculation behavior was also observed in the case of heavy mineral oil with 0.1% v/v SPAN 80, with

$Ca \sim 1.75 \times 10^{-4}$. Interestingly, for the heavy mineral oil with 0.1% v/v SPAN 80 system, the flattening of the droplet interfaces to form the thin film is significant enough that it can be visibly observed in the image, as shown in **Figure 4C**. At this instant, there is a slight step decrease in the center-to-center distance between droplets indicating significant deformation of the droplet surfaces in contact. Again, it is important to note that this is different from the upper critical capillary number occurring at high speeds, above which droplets will merely flow past the trapped droplet without the film draining completely.

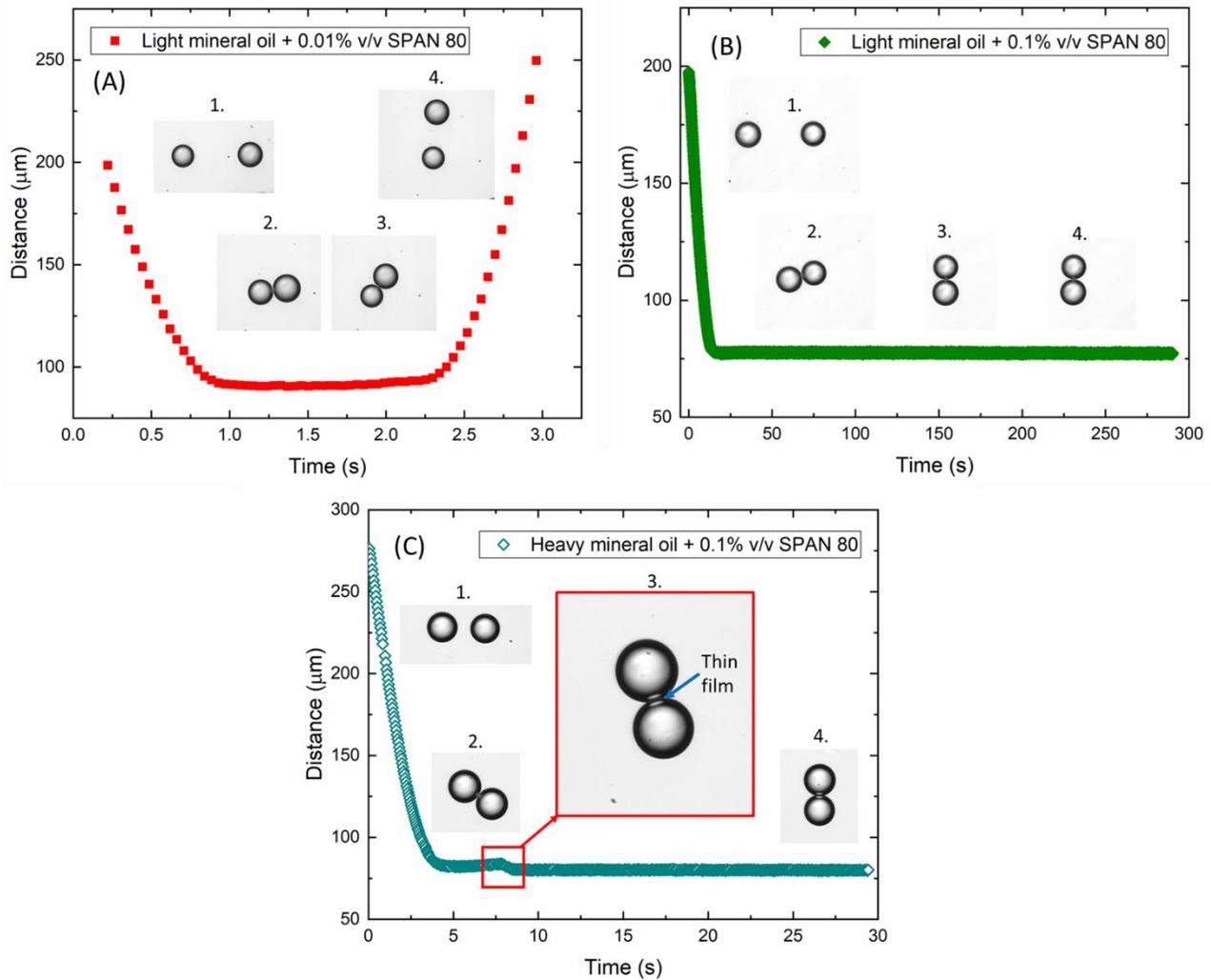

*Figure 4: Center-to-center distance between droplets as a function of time for (A) Light mineral oil + 0.01% v/v SPAN 80 system with $Ca \sim 1 \times 10^{-3}$, where droplets glance off each other without coalescing. The trapped and incoming droplet radii are 44 µm and 49 µm respectively (B) Light mineral oil + 0.1% v/v SPAN 80 system with $Ca \sim 1 \times 10^{-4}$ where droplets flocculate, but do not coalesce. The trapped and incoming droplet radii are 36 µm and 38 µm respectively. (C) Heavy mineral oil + 0.1% v/v SPAN 80 system with $Ca \sim 1.75 \times 10^{-4}$ where droplets flocculate, but do not coalesce. The trapped and incoming*

*droplet radii are 42 μm and 41 μm respectively. Inset figures 1-4 in all the above show droplet orientations at various instants of time. The zoomed-in inset figure 3 in (C) shows the thin film region formed between the flocculated droplets (see bright region between droplets).*

## 3.2 Drop rotation: head-on vs glancing collisions

The offset between the symmetry axis of the flow and the line of centers of the two droplets has been found previously to influence droplet coalescence in linear flow fields.[42] Ideally, two droplets would collide head-on, often leading to coalescence without rotation about each other. However, even the most infinitesimal offset causes the incoming droplet to rotate in the flow, and these collisions are termed glancing collisions. In the experiments in this study, almost all collisions are glancing collisions, because the finite time period between the moment when the first droplet is trapped and the incoming droplet collides with it may cause the trapped droplet to drift from the central axis of the microfluidic device. The angle between the compression axis of the flow and the line of centers of the two droplets is defined as the angle of approach ($\theta$) as indicated in the inset figure in **Figure 5A.** Additionally, $\theta_{ct}$ is defined as the angle at which the center-to-center distance between the droplets is $R_t + R_i$, where $R_t$ is the radius of the trapped droplet and $R_i$ is the radius of the incoming droplet, and $\theta_{co}$ is defined as the angle at which the droplets coalesce. In the planar extensional flow in the cross-slot device, the external force $F_{ext}$ acting along the line of centers in the undisturbed flow is known to vary according to the equation[42]

$$F_{ext}(t) = 2f(\lambda)\eta_c U R \cos\{2\theta(t)\}, \tag{1}$$

where $\lambda$ is the viscosity ratio, $U$ is the relative velocity at $4R_t$ (here, equal to the velocity of the incoming droplet) and $R = \frac{2R_t R_i}{R_t+R_i}$ is the geometric mean radius of the interacting droplets. As the incoming droplet moves from the compressional quadrant of the flow to the extensional quadrant, $\theta$ increases from 0° to 90° for the case where the droplets do not coalesce. On the other hand, for a head-on collision leading to coalescence, the rotation of the droplet remains small ($< 5°$) for the entire duration of droplet interaction as shown in **Figure 5A,C**. Therefore, the force $F_{ext}$ remains positive and constant for the entire duration of a head-on collision once the droplets have come into apparent contact with a thin film between them, and $F_{ext}$ is countered by the hydrodynamic lubrication force in the thin film, up to the point where the film thins sufficiently to cause rupture due to attractive van der Waals force.

On the other hand, for glancing collisions, the droplets rotate about each other in the flow and the angle of approach increases. Hence, $F_{ext} > 0$ in the compressional quadrant of the flow field where $\theta < 45°$, $F_{ext} = 0$ when $\theta = 45°$ and finally, $F_{ext} < 0$ when $\theta > 45°$ in the extensional quadrant of the flow field. Surprisingly, here we observe that for glancing collisions (or collisions with some initial offset between droplets), the droplets tend to coalesce not only in the compressional quadrant of the flow, but also in the extensional quadrant just after the force starts to become negative, so long as the capillary number is smaller than the critical upper value for coalescence. This means that the film has thinned sufficiently such that the attractive van der Waals force is dominant and can cause film rupture by overcoming the repulsive external and lubrication forces. **Figure 5B** shows the angle of coalescence $\theta_{co}$ as a function of the angle of contact $\theta_{ct}$ for the light mineral oil + 0.05% v/v SPAN 80 system. In general, the angle at which coalescence occurs increases with an increase in the offset or angle of contact, and it is observed that several coalescence events occur in the extensional quadrant of the flow. In other words, for glancing collisions, despite the external force starting to pull the droplets apart, the film still ruptures and the droplets coalesce. Additionally, **Figure 5C** shows that droplets may rotate up to 65°, before coalescence can occur, and this large degree of rotation occurs in droplets travelling at high velocities. The dashed line in **Figure 5C** indicates the transition from head-on to glancing collisions. Borrell et al.[42] found, via four-roll mill experiments, that the time history of the force and the evolution of the shape of the thin film play significant roles in the process. As noted previously, $F_{ext}$ is time-varying in a glancing collision. When the droplet enters the extensional quadrant of the flow, a suction force is created on the droplet leading to local deformation and dimpling of the thin film.[42,43] The effect may be enhanced even further in the presence of surfactants, and this dimpling may create an instability along the rim, leading to coalescence. The local film deformation dynamics may explain why droplets coalesce in the extensional quadrant of the flow when the external force starts to pull them apart.

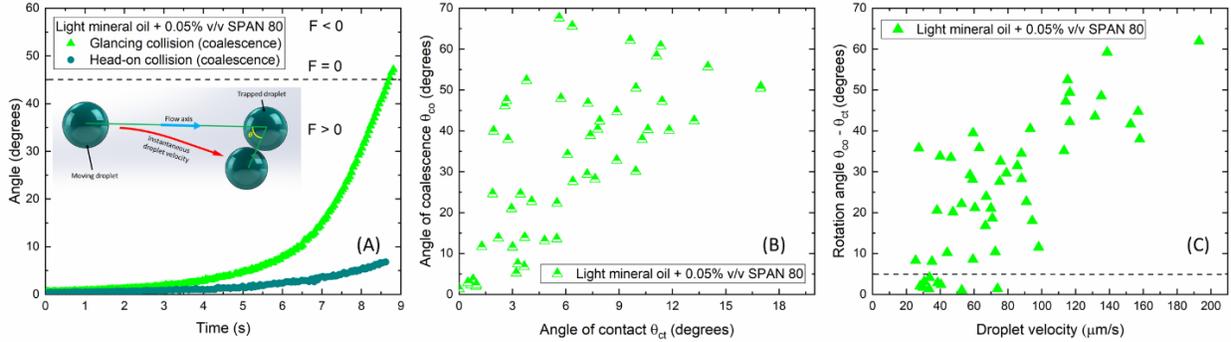

*Figure 5: (A) Angle of approach as a function of time showing head-on and glancing collisions. The system here is light mineral oil + 0.05% v/v SPAN 80. Both cases lead to coalescence. The inset figure shows two positions of the incoming droplet, with the angle of approach θ indicated. The dotted line indicates the angle (45°) at which the external force changes sign. (B) Angle at which droplets coalesce plotted against angle at which the droplets make contact (when center to center distance = sum of radii) for the light mineral oil + 0.05% v/v SPAN 80 system. (C) Degree of droplet rotation as a function of the velocity of the incoming droplet for the light mineral oil + 0.05% v/v SPAN 80 system. The dotted line indicates the transition from head-on to glancing collisions.*

*3.3 Film drainage time as a function of droplet size and velocity*

The hydrodynamic trap device in this work is used to study coalescence of surfactant-stabilized droplets over a range of droplet sizes and speeds. In this section, we consider the effect of droplet radius in comparison to the channel height on the film drainage time during coalescence. In the hydrodynamic trap, both monodisperse and polydisperse collisions may occur. The most common example is the case when two monodisperse droplets coalesce, following which the trapped droplet becomes larger than the incoming droplets and the subsequent coalescence events involve polydisperse droplets. In these cases, provided the incoming droplet speeds are not significantly different, the film drainage time $t_d$ increases for each consecutive coalescence event, such that as the trapped droplet at the center of the cross-slot becomes larger, the film drainage time increases slightly with each coalescence event. **Figure 6A** shows the film drainage time measured in three such serial coalescence events, for light mineral oil with two different surfactant concentrations (0.05% and 0.005% v/v SPAN 80) in the continuous phase. It is evident that with each consecutive coalescence event in the series, the film drainage time is higher, and the radius of the trapped droplet increases. Moreover, the film drainage time is significantly lower for the lower surfactant concentration. One possible explanation for the increase in film drainage time with increase in droplet radius is that as the radius of the trapped droplet increases, it may settle to the bottom of the channel since it is stationary and the droplet phase (water) is denser than the surrounding phase (oil). This settling may cause an offset in the vertical plane, in addition to any offset in the horizontal plane, which may lead to greater rotation between the droplets and delay coalescence. Moreover, an increase in the radius may lead to a larger film being formed, which may in turn also delay coalescence. Alternatively, confinement of the droplet in the channel may cause delayed coalescence as discussed in the following paragraphs.

**Figure 6B** shows the film drainage time as a function of trapped droplet radius for light and heavy mineral oils with 0.05% v/v SPAN 80 in the continuous phase. For the systems studied here, the film drainage time shows an increase with trapped droplet radius. The data shown here includes serial coalescence events as described above, as well as polydisperse collisions in general, and in both cases the radius of the trapped droplet ($R_t$) has been used. Both moderately confined droplets and highly confined droplets are studied in this work, with the confinement ratio $2R/h$ ranging from ~ 0.5 to 1.6, where $h$ is the channel height, here ~ 100 µm. In applications such as fuel filtration, droplets are squeezed through highly confined environments. Hence, it is important to understand the effect of confinement on the film drainage time during coalescence of both confined and unconfined droplets. Some studies have previously investigated the coalescence between droplets in microfluidic devices under confinement.[43,44] For instance, Chen et al. studied the coalescence of Newtonian droplets in a Newtonian matrix under shear in a parallel-plate device and found that confinement increases the likelihood of droplet coalescence while also increasing the rotation angle in flow at which coalescence occurs.[44] It was also shown that for highly confined droplets the force exerted by the walls in the tangential direction becomes equal to 10% of the hydrodynamic force, therefore significantly influencing the dynamics of drainage. Thus, in addition to stabilizing forces due to surfactants at the interface, the wall force introduces an additional resistance to film drainage that slows down coalescence. For highly confined droplets ($\frac{2R}{h} > 1$), the drop shape will be pancake-like as opposed to spherical. Previously, deformation of highly confined droplets in microfluidic channels has been analyzed as a 2D problem as opposed to a full 3D problem in the case of unconfined droplets.[45,46] Drawing an analogy to these studies, it is reasonable to infer that the film drainage in the case of highly confined droplets in microfluidic channels would be a 2D problem of thin film drainage, wherein the film drains predominantly from the sides of the droplet with fluid motion being restricted in the vertical direction, unlike the unconfined case, wherein drainage is possible in all directions. Moreover, large droplet size leads to the creation of a larger thin film region where the droplets are in contact. Since the maximum pressure differential responsible for driving flow out of the thin film scales as the capillary pressure which is $O(\frac{2\gamma}{r})$, where $r$ is the thin film radius, larger films lead to smaller pressure gradients, resulting in longer film drainage times. From our observations in this experiment, two main conclusions can be drawn about the effect of droplet size on film drainage time – 1) increasing droplet diameter, and therefore confinement, increases the film drainage time and 2) confinement tends to hold droplets in contact for a longer period of time by slowing down relative motion, and therefore increases the probability of coalescence.

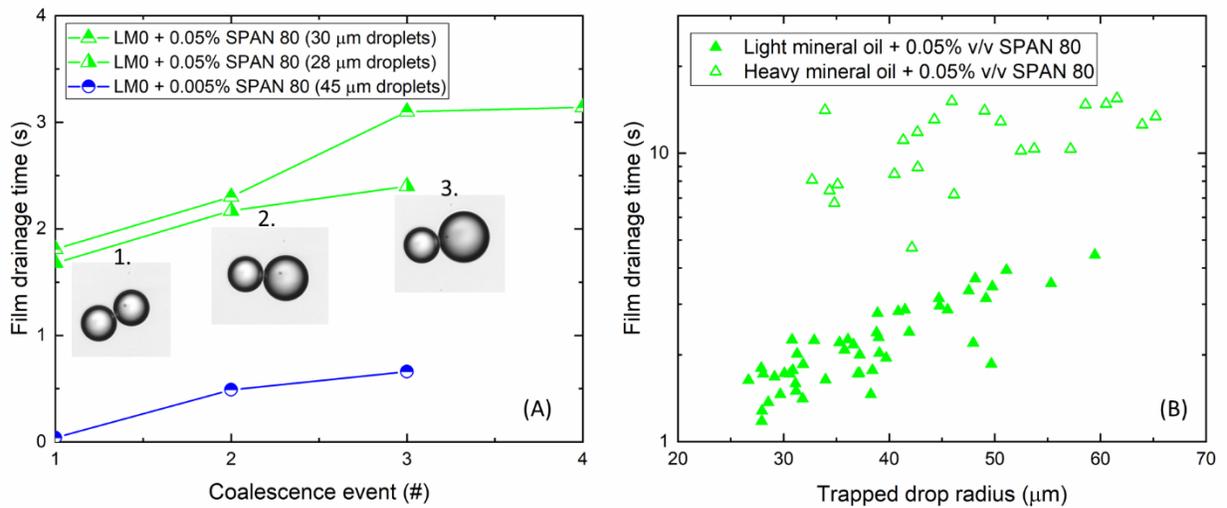

*Figure 6: (A) Film drainage time for a series of coalescence events with a single trapped center droplet, which grows with each coalescence event. Different symbols indicate different sizes of incoming droplets. Two different concentrations of SPAN 80 in light mineral oil are shown – 0.05% and 0.005% v/v. Inset images 1-3 show relative droplet sizes for the serial coalescence of 28 μm droplets for the 0.05% v/v SPAN 80 system. (B) Film drainage time as a function of trapped droplet radius for light mineral oil + 0.05% v/v SPAN 80 (closed triangles) and heavy mineral oil + 0.05% v/v SPAN 80 (open triangles).*

Next, we consider the relative importance of viscous and interfacial forces in controlling the coalescence and, in particular, the film drainage time. The capillary number defined previously as $Ca = \frac{\eta_c U}{\gamma}$ is used to characterize the coalescence behavior in the absence of inertial effects. The relative velocity $U$ between the droplets is measured at a distance of $4R_t$ from the trapped droplet. Gentle collisions which ensure longer contact times between droplets favor coalescence. Consider that the incoming droplet is moving with a fluid having a static pressure $p$, which approaches the trapped droplet with a relative velocity $U$. As the droplets come into close proximity, the lubrication pressure in the thin film between the droplets increases, with the total pressure becoming $p + \frac{1}{2}\rho_c U^2 + \frac{2\gamma}{r}$, where $\rho_c$ is the density of the outer fluid, $\frac{2\gamma}{r}$ is the capillary pressure and $r$ is the radial extent of the thin film. Thus, the difference between this total pressure within the thin film and the static pressure $p$ in the bulk fluid drives flow of the continuous phase out of the film. An increase in the equilibrium interfacial tension $\gamma$ would lead to an increase in the pressure gradient, leading to short film drainage times. Depending on the relative magnitudes of $\frac{1}{2}\rho_c U^2$ and $\frac{2\gamma}{r}$, increasing the velocity of the incoming droplet can have two effects. It can either lead to an increase in the driving pressure gradient, or it can cause significant flattening of the film, leading to a local increase in the radius of the film $r$, leading to a smaller pressure gradient and longer drainage times. At high $Ca$ in the range ~$10^{-3}$ - $10^{-2}$ and higher, Yang et al.[47] found via four-roll mill experiments that the drainage time did not exhibit a dependence on $Ca$ when $Ca < \sim 10^{-3}$ although in their work, the majority of experiments were performed at $Ca > \sim 10^{-3}$. In the experiments here, $Ca < \sim 10^{-3}$ for most cases.

The fluid-fluid interfaces in this study are partially mobile at high surfactant concentrations and might even be fully mobile as surfactant concentration is decreased (see discussion in Section 3.4).

Therefore, the outer problem or the flow pushing the droplets together may have a significant effect on the film drainage time.[48–50] **Figure 7A and B** show the film drainage time $t_d$ divided by the geometric mean droplet radius $R$ (since $t_d$ increases with $R$) as a function of the velocity $U$ for the light and heavy mineral oil systems across the entire range of surfactant concentrations. At a given droplet radius and for a given surfactant concentration, the film drainage time shows a weakly decreasing trend with an increase in velocity. As the velocity of the incoming droplet is increased, the external force $F_{ext}$ pushing the droplets together increases, and this leads to a decrease in the film drainage time. Moreover, for the lowest surfactant concentrations, the film drainage time essentially shows no dependence on the droplet velocity. This observation is consistent with the observations by Zhou et al,[14] who also observed a decrease in film drainage time with an increase in velocity in a microfluidic droplet collision experiment in a cross-slot with nanoparticle-coated droplets. Moreover, they also observed that for droplets not stabilized by nanoparticles, there was no dependence of drainage time on droplet velocities.

**Figure 7C and D** show the non-dimensional film drainage time plotted as a function of the capillary number $Ca$. A certain degree of collapse onto a single curve is achieved for the higher SPAN 80 concentrations with light mineral oil, but the 0.005% v/v curve is essentially flat and hence does not collapse onto the other curves. With heavy mineral oil in the continuous phase, better collapse was obtained. However, the dependence of the dimensionless drainage time on $Ca$ is weakly decreasing at best. This observation is corroborated by previous work by Frostad et al[21] using a cantilevered capillary force apparatus (CCFA), where the total contact time before coalescence decreased with increasing $Ca$ over the same range of $Ca$ as the present study. In the present study, however, the upper limit on $Ca$ is $\sim 2 \times 10^{-3}$, and the range is limited by the conditions required for stable droplet trapping. Moreover, as noted previously, the alteration of the flow field around confined droplets likely has an influence on the scaling for film drainage time and the relationship is convoluted by the different rotation angles before coalescence.

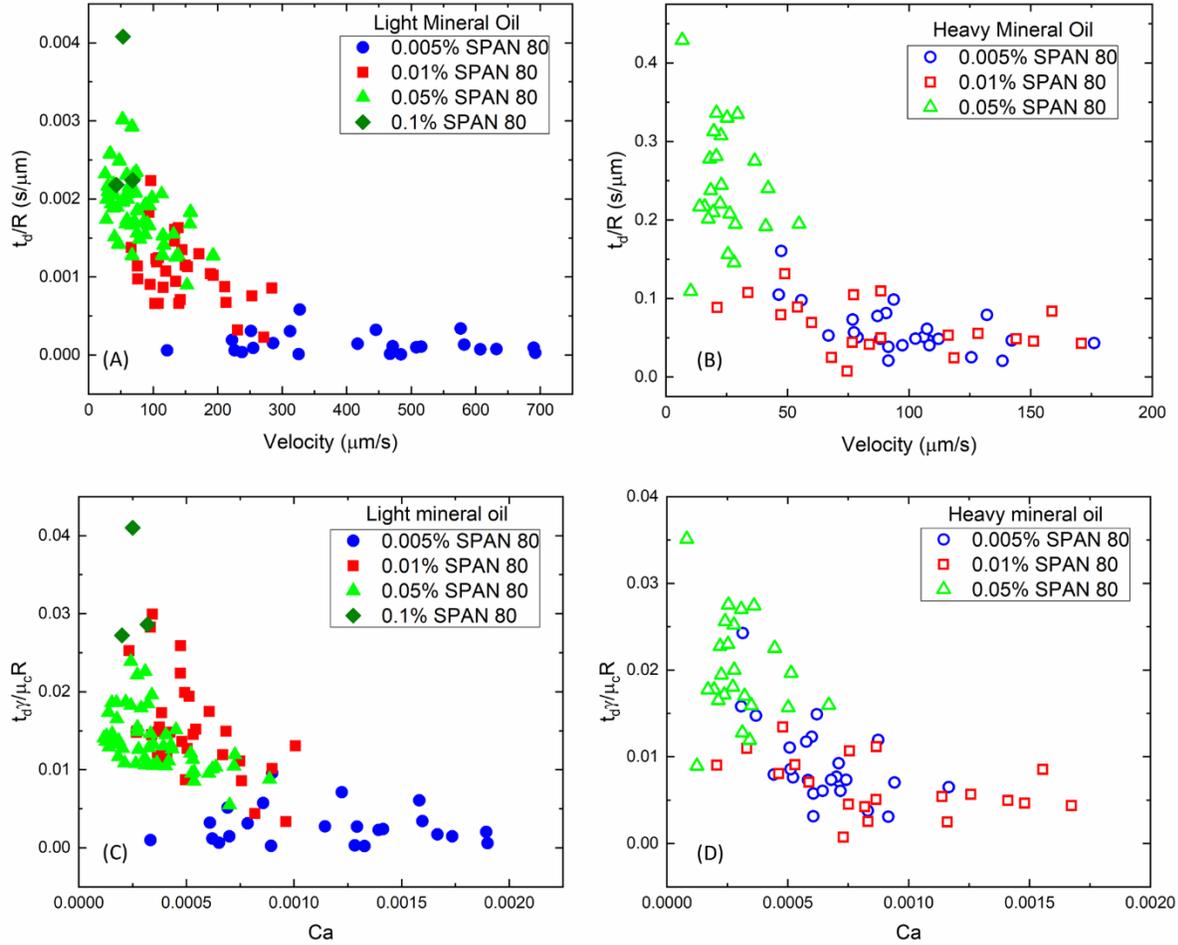

*Figure 7: (A) Film drainage time normalized by radius, as a function of droplet velocity at various concentrations of SPAN 80 in light mineral oil (B) Film drainage time normalized by radius, as a function of droplet velocity at various concentrations of SPAN 80 in heavy mineral oil (C) Dimensionless film drainage time as a function of capillary number at various concentrations of SPAN 80 in light mineral oil (D) Dimensionless film drainage time as a function of capillary number at various concentrations of SPAN 80 in heavy mineral oil*

*3.4 Effect of continuous phase viscosity, surfactant concentration and gradients*

The concept of interfacial mobility and its correlation to film drainage was introduced in Section 3.1. Briefly, high viscosity ratios ($\lambda = \frac{\eta_d}{\eta_c}$) and high surfactant concentrations give rise to immobile or no-slip interfaces, while low viscosity ratios and low surfactant concentrations give rise to mobile or no-stress interfaces. Therefore, the relative viscosity of the dispersed and continuous phases is a key factor in determining the mobility of the liquid-liquid interface. The water-in-light and heavy mineral oil systems studied here have viscosity ratios equal to 0.037 and 0.008 respectively, with the continuous phase (mineral oil) being orders of magnitude more viscous than the dispersed phase (water). However, the systems studied here also contain surfactants, and surfactants are known to reduce interfacial mobility. While this effect may be significant at concentrations above CMC, most concentrations here are at or below the CMC. Hence, it is reasonable to assume that the interfaces in this study are at least partially if not fully mobile, with a no-stress condition occurring at the interface, except at concentrations at or above CMC. It is also reasonable then to assume that the viscosity of the continuous phase controls the film drainage process.[25] **Figure 8A** shows the average film drainage times for the all the systems studied here and across the entire range of surfactant concentrations. Clearly, the film drainage time is longer with heavy mineral oil in the continuous phase than with light mineral oil in the continuous phase. When the dispersed phase viscosity is small, Chesters[25] postulated that the drainage is controlled predominantly by the resistance offered by the thin film to deformation, and the film thickness $h$ decays as

$$h = h_0 \exp\left(-\frac{t}{\tau}\right), \qquad (2)$$

where $h_0$ is the initial thickness of the film and $\tau$ is a characteristic time given by $\tau = \frac{3\eta_c R}{2\gamma}$. Assuming that the film ruptures when $h$ reaches a critical thickness $h_c$, the above equation can be rearranged to give the film drainage time $t_d$ for a fully mobile film as

$$t_d = \frac{3\eta_c R}{2\gamma} \ln\left(\frac{h_0}{h_c}\right). \qquad (3)$$

Thus, the film drainage time should increase when the continuous phase viscosity increases. The initial film thickness $h_0$ depends largely on the strength of the flow field, and in a weak flow field with low $Ca$, the drops will remain spherical. On the other hand, at high $Ca$ strong interfacial deformation arises with pressures exceeding the Laplace pressure, and the drop deformation delays coalescence. Larger $Ca$ results both from higher incoming droplet velocity and higher continuous phase viscosity, such that with the heavy mineral oil systems, both the interfacial deformation and the resistance of the film to drainage are higher, resulting in longer film drainage times than the systems with light mineral oil in the continuous phase. Finally, note that while Equation (3) provides a good estimation of drainage time for pure fluids, surfactant-laden interfaces may behave differently and further resist drainage due to repulsive forces and surfactant redistribution along the interface, which will be discussed in the following paragraphs.

Another trend that emerges from **Figure 8A** is unsurprising – as the SPAN 80 concentration is increased, the film drainage time increases with both light and heavy mineral oils in the continuous phase. The slower film drainage is, of course, attributable to the surfactant-induced enhanced stability of the liquid-liquid interface and a decrease in the interfacial tension according to equation (3) and **Figure 2**. One of the main mechanisms by which surfactants stabilize interfaces is by occupying sites at the interface to increase the surface pressure and lower the interfacial tension, which in turn results in a smaller thermodynamic driving force for coalescence. As seen in **Figure 2**, the interfacial tension can be as low as 5 mN/m for surfactant-laden mineral oil-water interfaces. For the lowest surfactant concentrations studied in light mineral oil (0.005% v/v SPAN 80), the interfacial tension is ~ 10 mN/m, and film drainage times are rapid, resulting in larger spread in the measured film drainage times due to limitations on imaging frame rate. All the droplets brought into contact in the trap region coalesce at this low surfactant concentration, indicating extremely low emulsion stability. On the other hand, at the highest surfactant concentration, here 0.1% v/v SPAN 80, very few coalescence events are observed, and film drainage times are significantly longer. As noted in Section 3.1, droplets may flocculate at extremely low $Ca$ (as opposed to being advected away at high $Ca$), and this flocculation behavior is due to a balance between flow-induced and surfactant-induced stabilizing forces in the extensional quadrant of the flow and attractive van der Waals forces between the droplets. This flocculation is only observed at the highest surfactant concentration of 0.1% v/v SPAN 80, which is two times the critical micelle concentration. Some studies have shown that for surfactant concentrations above the CMC, addition of more surfactant does not have a significant impact on drop size distribution during emulsion formation.[51] However, it is also known that inverse micelles can induce depletion interactions in the thin film.[52,53] Experimental findings suggest that when a structuring colloid, such as micelles, is present in the thin film, at sufficiently small separations they may give rise to oscillatory structural forces within the thin film.[54] While it is possible that some or all of these interactions lead to the long drainage times and flocculation behavior observed in our experiments, the precise role of micelles in stabilizing the film against drainage is a subject of current research in our laboratory. Finally, the dilatational elasticity or viscosity of the adsorbed surfactant layer may contribute to enhanced emulsion stability at high concentrations.[55] However, Santini et al.[56] measured the dilatational elasticity and viscosity of SPAN 80 at the paraffin oil-water interface over a range of frequencies from 0.01 Hz – 20 Hz, and found no significant correlation between emulsion stability and dilatational properties. In fact, at the high frequency limit, which is thought to control coalescence, and at a concentration 50 times the CMC, a decrease in dilatational elasticity was observed, which the authors attributed to the presence of inverse micelles.

In addition to the flocculation observed at 0.1% v/v SPAN 80 in light mineral oil, another interesting phenomenon was observed when a third droplet moves into the trap region and makes contact with the flocculated droplets. As shown in **Figure 8B – 1**, droplet 1 is trapped at the center of the cross-slot, while droplet 2 is the incoming droplet. Due to low capillary number ($Ca \sim 1 \times 10^{-4}$), droplets 1 and 2 flocculate. Subsequently, droplet 3 makes contact with droplet 1 in a glancing collision and rotates up to 83° into the extensional flow quadrant, but instead of flocculating, it coalesces with droplet 1 as seen in **Figure 8B – 5**. Thus, even though $Ca$ is the same for both collision events, the latter led to coalescence, with droplet 3 rotating almost up to 90° in the extensional quadrant of the flow. Note that although interesting, this delayed coalescence is quite rare, and the conditions for initial flocculation (low $Ca$, high surfactant concentration) must be met for such an event to occur. The film drainage time for this coalescence event was measured to be 19.59 seconds, which is significantly longer than the other coalescence events for the same system. Similar flocculation behavior of two droplets followed by coalescence of the third droplet was also observed in the case of heavy mineral oil with 0.1% v/v SPAN 80. Coalescence did not occur for any other collision events in this system; instead, droplets flocculated at low $Ca$. When coalescence of a third droplet was observed following flocculation of the first two, the capillary number was $\sim 1.5 \times 10^{-5}$. In the following paragraphs, we attempt to explain this coalescence behavior by breaking down film drainage into three sub-stages.

One of the most comprehensive descriptions of the role of surfactants in film drainage, or the inner problem, has been provided by Hudson et al.,[9] who conducted coalescence experiments of droplets in a shear cell. The Peclét number ($Pe = GR^2/D$) is used to characterize the importance of advection vs diffusion, with $D$ being the diffusivity of the surfactant in the bulk fluid phase, $R$ the droplet radius and $G$ being the strain rate. At the interface, surfactant transport is much more complicated, and it is generally agreed upon that the motion of the surfactant along the interface during film drainage may have a significant impact on the drainage time. To account for interfacial motion of surfactant molecules, Hudson et al.[9] proposed the use of an interfacial Peclét number ($Pe_\gamma = GR^2/D_\gamma$), where $D_\gamma$ is the surfactant diffusivity at the interface, in combination with a Marangoni number ($Ma = \frac{\Pi}{\gamma_0 Ca}$), where $\Pi = \gamma_0 - \gamma$ is the surface pressure and $\gamma_0$ is the initial interfacial tension of the clean interface, to fully describe the surfactant behavior. The Peclét describes the relative importance of convective to diffusive motion of surfactants at the interface, whereas the Marangoni number indicates the relative importance of surface tension gradient-induced motion to convective motion at the interface. The authors postulated that the film mobility does not remain the same for the duration of the film drainage stage but evolves based on the values of $Pe_\gamma$ and $Ma$, resulting in three sub-stages of film drainage.

For the light mineral oil systems studied here, the Marangoni number ($Ma$) can be estimated as $Ma = \frac{\Pi}{\gamma_0 Ca} \sim 1.58 \times 10^3$ for 0.005% v/v SPAN 80 and $Ma \sim 1.77 \times 10^3$ for 0.1% v/v SPAN 80 assuming $Ca = 5 \times 10^{-4}$. Also, the interfacial Peclét number $Pe_\gamma = \frac{GR^2}{D_\gamma} \sim 20.25 \gg 1$, assuming a typical strain rate $G = 0.5\ s^{-1}$ and the interfacial diffusivity of SPAN 80 to be $D_\gamma = 5 \times 10^{-11}\ m^2/s$.[57] Based on these calculations, and the large value of the interfacial Peclét number, it can be inferred that convective effects will be dominant at the interface, giving rise to large interfacial concentration gradients and consequently large Marangoni stress at the interface, shown schematically in **Figure 3B**. However, note that the above equations are only valid for insoluble surfactants, and a detailed analysis for soluble surfactants would require knowledge of the adsorption/ desorption rate constants. Moreover, with soluble surfactants, the diffusion timescale is significantly shorter for small, micrometer-sized droplets and the transport is dominated by adsorption kinetics. Nevertheless, it is reasonable to assume that Marangoni stresses play a role in dictating the interfacial dynamics in our experiments, particularly at high surfactant concentrations.

Hudson et al.[9] elaborated on the role of surfactants and interfacial mobility on film drainage by breaking down the film drainage process in three stages – the first is a "mobile drainage" stage, where the fluid in the film drains and surfactant is swept outwards towards the rim of the thin film. In this stage, there exists a no-stress condition at the interface. The second stage is called a "transitional pause", where Marangoni stresses along the interface sweep the surfactant into the thin film region. In this stage, a dimple may form in the film, and the interface is immobilized due to Marangoni stresses. Large values of $Pe_\gamma$ would lead to a longer transitional pause and immobilize the interface for a longer period of time, whereas large values of $Ma$ lead to the transitional pause occurring earlier in the film drainage process. Finally, the interface is remobilized in the third stage, due to re-equilibration of the surfactant concentration, and the film drains rapidly leading to rupture along the rim. Thus, according to the explanations provided by Hudson[9] and Chesters and Bazhlekov[26], the mobility of the thin film evolves during the film drainage process, and depends largely on the surfactant and fluid properties.

The flocculation and delayed coalescence behavior shown in **Figure 8B** suggest that Marangoni effects may be significant in the droplet coalescence process at high enough concentrations, leading to surfactants being swept into the thin film region. Thus, the surfactant concentration is low on the side of droplet 1 that is not in contact with droplet 2, which locally increases the interfacial tension and leads to coalescence with droplet 3. Simulations of copolymer surfactant at a liquid-liquid interface by Dai et al.[58] showed that during film drainage, surfactant gets swept from the remainder of the droplet towards the thin film region, thereby locally increasing the concentration in the region surrounding the thin film. Of course, the timescale over which the interfacial surfactant concentration can re-equilibrate will dictate whether such delayed coalescence events will occur. During the transitional pause stage of film drainage, the interface is rendered immobile due to Marangoni stresses. It is likely that if the third droplet contacts the flocculated droplets during this transitional pause, it will coalesce due to the surfactant being swept into the thin film region. If the third droplet approaches the flocculated droplets during either the mobile drainage or remobilization stages of film drainage, it may not coalesce because the surfactant distribution is re-equilibrated at that time.

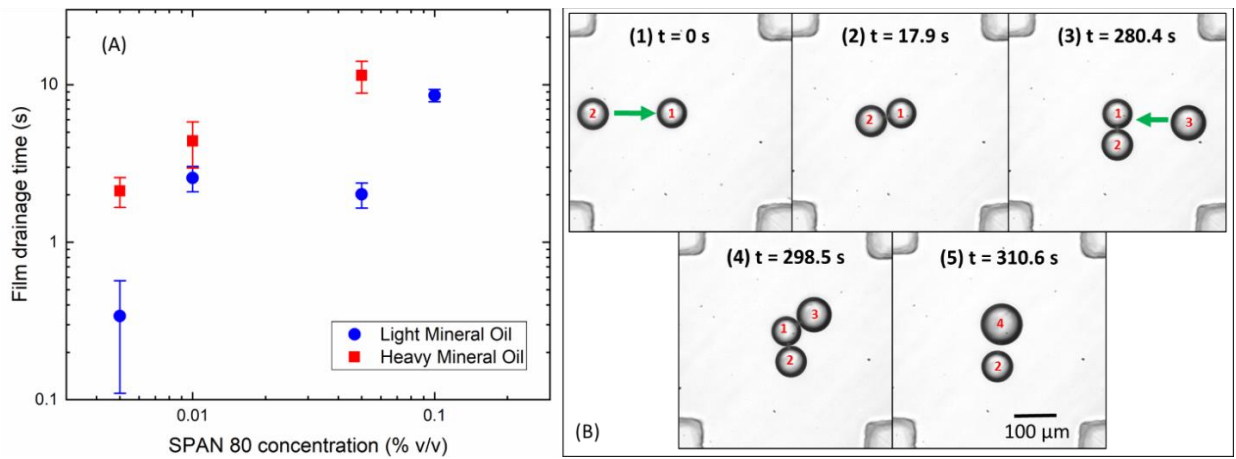

*Figure 8: (A) Film drainage time as a function of SPAN 80 concentration with light (blue circles) and heavy (red squares) mineral oils in the continuous phase. (B) Marangoni effects in the light mineral oil + 0.1% v/v SPAN 80 system showing 1. Approach of the first and second droplets, 2. Flocculation of droplets, 3. Approach of a third droplet 4. Rotation of the third droplet in flow and 5. Coalescence of the first and third droplets. Time stamps are indicated in the figure.*

**Conclusions**

A microfluidic hydrodynamic Stokes trap with a four-channel cross-slot design has been employed in this work to study the coalescence behavior of dispersed aqueous droplets formed in a continuous phase of mineral oil with SPAN 80 as a surfactant. Over a range of surfactant concentrations, droplets were formed at a T-junction geometry upstream of the cross-slot, and collisions between droplets leading to coalescence allowed measurement of film drainage time. As expected, film drainage time increases with increase in continuous phase viscosity, surfactant concentration and droplet size. Interesting coalescence dynamics emerge with surfactant-laden systems, which provide fundamental insights into the role of surfactants in film drainage. At extremely high Capillary numbers, insufficient contact time between droplets results in the droplets being carried away by the extensional flow before coalescence can occur despite the droplets being in contact in the cross-slot. Increase in droplet size and therefore droplet confinement in the microfluidic channel leads to longer film drainage times and greater probability of coalescence. This finding is important to emulsion separation processes such as filtration, where droplets are often squeezed through confined spaces. Further, at high surfactant concentrations (1000 ppmbv or 0.1% v/v) and low capillary numbers, droplets tend to flocculate in the trap region instead of coalescing, indicating a balance of attractive van der Waals force and repulsive force due to surfactant and extensional flow in the cross-slot device. Moreover, the importance of surfactant redistribution at the interface and the resultant Marangoni stresses becomes apparent when delayed coalescence of a third incoming droplet with the flocculated droplets occurs. The observations support the existing theory that the film mobility is not constant during film drainage. Instead, it evolves during the coalescence process due to surfactant concentration gradients, and results in immobilization and dimpling of the thin film, both of which suppress coalescence of droplets. The observations from this work open up avenues for further exploration of controlled coalescence of microscale droplets relevant to various emulsion separation applications of commercial importance and provide important fundamental understanding about the role of surfactants in droplet coalescence.

**Acknowledgements**

The authors would like to thank Professor Charles Schroeder, Dr. Anish Shenoy and Mr. Dinesh Kumar from the University of Illinois at Urbana-Champaign for their help in setting up the Stokes trap. The authors would also like to thank Dr. David Giles at the Polymer Characterization Facility and Dr. Wieslaw Suszynski at the Coating Process Fundamentals Lab, both in the Department of Chemical Engineering and Materials Science at the University of Minnesota. This work was primarily funded by Donaldson Company (Bloomington, MN), including support for S.N., and carried out in the Department of Mechanical Engineering at the University of Minnesota Twin Cities, College of Science and Engineering. This material is also based upon work supported in part by the National Science Foundation under NSF CAREER Grant No. 1554936, including partial support of I.M. Portions of this work were conducted in the Minnesota Nano Center, which is supported by the National Science Foundation through the National Nano Coordinated Infrastructure Network (NNCI) under Award Number ECCS-1542202. None of the authors have recent, present, or anticipated financial gain from this work, and there are no conflicts of interest to declare.